%% file: main.tex
\documentclass[reprint,
amsmath,amssymb,
aps,
longbibliography,
prl,
floatfix,
]{revtex4-1}

\input{pream}

\begin{document}

\title{Hydrodynamic interactions destroy motility-induced phase separation in active suspensions}

\author{Tingtao Zhou}
\author{John F. Brady}
\affiliation{Divisions of Chemistry and Chemical Engineering \& Engineering and Applied Science,
   California Institute of Technology
}

\date{\today}

\begin{abstract}
Motility-Induced Phase Separation (MIPS) is a distinctive phenomenon in active matter that arises from its inherent non-equilibrium nature. Despite recent progress in understanding MIPS in dry active systems, it has been debated whether MIPS can be observed in wet systems in which fluid-mediated hydrodynamic interactions (HI) are present.
We use theory and large-scale {\it Active Fast Stokesian Dynamics} simulations of the so-called squirmer model to show that collision-induced pusher force dipoles, which are present even for the simplest neutral squirmers (stealth swimmers), destroy MIPS when HI are included.
Both rotational and translational HI independently suppress phase separation: rotation by shortening a swimmer's persistence length (and thus reducing the swim pressure), translation by a confinement-scale advective fluid  flow.
We further clarify that collisional dipoles between swimmers and boundaries can generate attractive flows that promote particle aggregation observed in some previous simulations and experiments.
Finally, we show how to recover MIPS in fluidic environments by tuning the magnitude of the HI through brush-like surface coatings on the active particles.
\end{abstract}

\maketitle

Active matter exhibits rich and fascinating collective behaviors, such as flocking, bacteria turbulence~\cite{dunkel2013fluid,dombrowski2004self} and motility-induced phase separation (MIPS)~\cite{cates2015motility}, opening a vast avenue for developing non-equilibrium statistical mechanics. Among these, MIPS -- the coexistence of dense and dilute phases of active particles at sufficiently high denisty and persistence --- is theoretically predicted for simple self-propelling particles with merely excluded volume interactions, such as Active Brownian Particles (ABPs). 
Phenomenologically, it has been observed that particles in higher-density regions slow down and self-trap to form a dense phase in coesistence with a dilute phase. This insight has been  developed into field-theory descriptions for MIPS~\cite{tailleur2008statistical,fily2012athermal,speck2014effective}. 
An alternate mechanical perspective shows that MIPS is a non-equilibrium phase separation governed by a mechanical force balance between the two phases~\cite{takatori2014swim,omar2023mechanical}.

Despite theoretical progress in understanding dry MIPS, such idealized conditions are often complicated by other physics in reality:
for microbes or active colloidal particles/droplets, the surrounding fluid mediates long-ranged, many-body hydrodynamic interactions (HI) that may alter the collective behavior. The simplest model to investigate the role of HI 
on active matter is the so-called squirmer model~\cite{lighthill1952squirming,blake1971spherical}. 
A squirmer is a force- and torque-free sphere with a prescribed surface slip velocity 
due to active swimming at low Reynolds number.  In addition to translating 
without disturbing the fluid, an individual squirmer can also generate a force dipole, or stresslet, flow field that decays as $1/r^2$; a positive dipole is a `pusher' or extensile swimmer, while a negative dipole is a `puller' or contractile swimmer.  A zero force dipole 
is a `neutral' swimmer, with the velocity disturbance decaying as $1/r^3$, the form of the velocity disturbance caused by phoretic motion.  In dilute suspensions, HI can be approximated by far-field multipole expansions, whereas in densely packed suspensions,  near-field lubrication interactions dominate.  The full range of HI is  required to describe suspensions at intermediate concentrations.


On one hand, far-field HI predict hydrodynamic instabilities and strong mixing in a dilute suspension of pushers, while a suspension of pullers remains stable~\cite{saintillan2008instabilities}. 
On the other hand, numerical studies with full HI highlight the importance of lubrication interactions in pattern formation, such as aggregation and band formation for a monolayer of `pullers'~\cite{ishikawa2008coherent,ishikawa2008development} and bound states for a neutral squirmer {\it Volvox}~\cite{drescher2009dancing,ishikawa2020stability}. 
Regarding MIPS, a `squirming disks' model~\cite{matas2014hydrodynamic} has suggested suppression of MIPS due to the rotational diffusion of active disks enhanced by near-field HI.  
However, another simulation~\cite{zottl2014hydrodynamics} of squirmers confined in a quasi-2D channel, but rotating and translating in 3D, has shown clustering.
Given the small number of particles ($O(200)$) used in previous numerical simulations, an investigation of larger systems is needed to resolve the seemingly inconsistent observations and to reveal  how HI affects MIPS.

\begin{figure}[t]
\centering
\includegraphics[width=0.99\columnwidth]{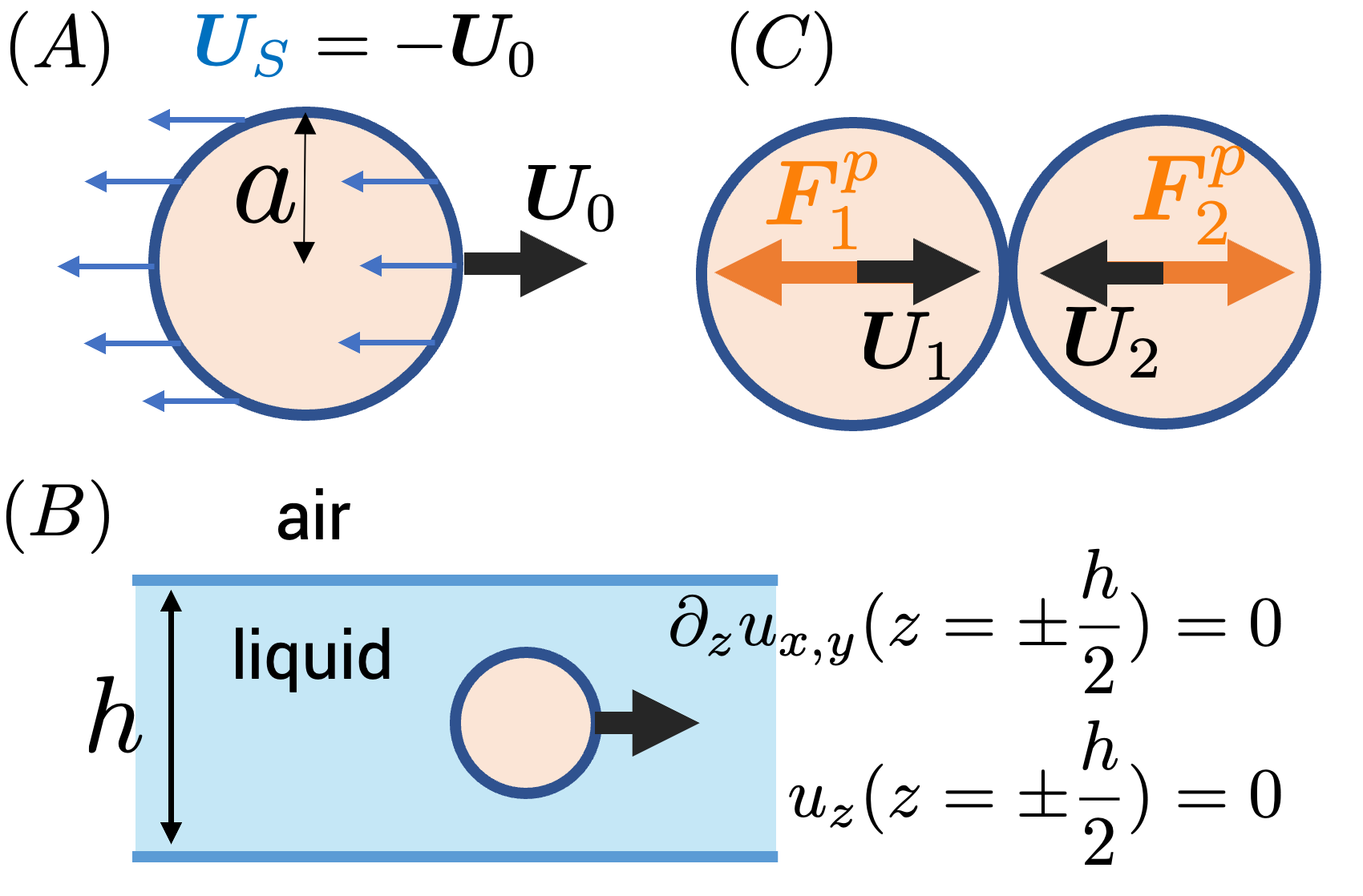}
\centering
\caption{
(A) Sketch of a neutral stealth swimmer. The surface slip velocity $U_S$ is a constant, hence no disturbance flow when swimming alone.
(B)  A swimmer confined in a liquid film of thickness $h$. At the liquid-air interfaces, zero shear-stress boundary conditions apply. The swimmer's motions are constrained in 2D. (C) Sketch of a collision-induced dipole. During a head-on collision, the hard-sphere repulsion between particles 1 and 2 induces a `pusher'-like dipolar flow field at leading order. 
\label{fig:fig1}}
\end{figure}

In this Letter, we combine theory and large-scale simulations to show that full HI destroy MIPS due to a many-body effect---collision-induced dipoles. Using Active Brownian Particles (ABPs) as a basic model for dry active matter, we focus on the simplest swimmer, a so-called stealth swimmer, as schematically shown in Fig.~\ref{fig:fig1}(A), to demonstrate the leading-order effects of HI. A stealth swimmer generates no velocity disturbance as it moves through the fluid, but when two particles `collide' the non-hydrodynamic interparticle forces that prevent the swimmers from overlapping now create a pusher hydrodynamic velocity disturbance (Fig.~\ref{fig:fig1}(C)).  This collision-induced dipolar flow dominates the long-range HI and alone can destroy MIPS. 

To illuminate the role of boundary confinement, we focus on shear-stress-free and no-flux boundary conditions at $z=\pm h/2$ that represent liquid films, as shown in Fig.~\ref{fig:fig1}(B). 
As noted in previous studies on HI-induced like-charge attraction~\cite{squires2000like} and aggregation of bottom heavy swimmers~\cite{drescher2009dancing}, the collision between particles and a wall generates strong flows and fluid-mediated interactions between particles. To separate this mechanism, we constrain  the motion and orientation of the swimmers within the $xy$ plane.
Our simulation method extends a fast version of the Active Stokesian Dynamics method~\cite{brady1988stokesian,fiore2019fast,elfring2022active} to enable large-scale systems of $O(10^4)$ particles. 

{\it Numerical method---}To simulate a suspension of swimmers with HI, we integrate the dynamics of $N$ swimmers of radii $a$ according to the overdamped force/torque balance~\cite{elfring2022active}
\begin{equation}
0 = - \boldsymbol{\mathcal{R}}_{\cal FU}\cdot(\boldsymbol{\mathcal{U}} + \boldsymbol{\mathcal{U}}^s) +  \boldsymbol{\mathcal{F}}^p - \boldsymbol{\mathcal{R}}_{\cal FE}:\boldsymbol{\mathcal{E}}^s\, .
\label{eqn:EOM}
\end{equation}
In Eq.~(\ref{eqn:EOM}), $\boldsymbol{\mathcal{U}} = (\bU_1, \bU_2 \ldots \bU_N; \bOmega_1, \bOmega_2, \ldots \bOmega_N)$ is a vector of the translational and rotational velocities of all $N$ particles, and $\boldsymbol{\mathcal{U}}^s$ is the corresponding swim velocities.     
The grand (many-body) resistance tensors $\boldsymbol{\mathcal{R}}_{\cal FU}$ and $\boldsymbol{\mathcal{R}}_{\cal FE}$ couple the generalized force/torque to the velocity and rate of strain $\boldsymbol{\mathcal{E}}^s$ due to swimming, respectively, and are functions of the configuration of all $N$ particles.  
The nonhydrodynamic force $\boldsymbol{\mathcal{F}}^p$ is a short-range hard-sphere force that prevents particles from overlapping. 
The swim velocity of particle $\alpha$ relates to the squirmer coefficient $B_1$ by $\bU_\alpha^s = - U_0 \bq_\alpha=-\frac{2}{3}B_1\bq_\alpha$, with constant amplitude $U_0$ and direction $\bq_\alpha$.   
We ignore Brownian motion of the swimmers and assume that their reorientation comes from the inherent noise of the surface slip velocity ($\bOmega^s_\alpha$) that does not generate a disturbance flow, which is possible for spherical swimmers, and model this noise as a random process with rotational diffusivity $D_R$.  The symmetric first moment of activity, $\boldsymbol{\mathcal{E}}^s$,  relates to the squirmer coefficient $B_2$ by $a \bE_\alpha^s=-\frac{3}{5}B_2 \overline{\bq_\alpha\bq_\alpha}$, where the overbar denotes an irreducible tensor.  A negative $\beta=B_2/B_1$ is a pusher and a positive $\beta$ a puller; $B_2 =0$ is a neutral squirmer.  Eq.~(\ref{eqn:EOM}) is solved for $\boldsymbol{\mathcal{U}}$ and then the positions and orientations of a particle are updated according to: $d \boldsymbol{X}_\alpha/dt = \bU_\alpha$ and $d\bq_\alpha/dt = \bOmega_\alpha \times \bq_\alpha$, respectively.

Planar bounding surfaces, either no-slip or slip or stress-free, can be represented by image systems~\cite{lorentz1907allgemeiner}.  However, there is a very simple way for no-shear stress boundaries:  the 3D unit cell consists of spherical particles in the $xy$-plane and this plane is periodically replicated in the $z$-direction at a separation $h$.  In this 3D unit cell HI among particles are fully three dimensional (thus circumventing the 2D Stokes paradox), but by symmetry the motion of the particles are confined to the 2D  $xy$-plane with rotation about the $z$-axis.   Periodic boundary conditions are imposed in the $xy$-directions. When $h \rightarrow 2a$ the spherical particles are touching and form a 2D rod along the $z$-axis, while for $h \gg a$ we have a monolayer or film of particles immersed in an infinite fluid.

The simulations are initialized with particles at random positions with random orientations. 
Each simulation is run  for dimensionless time $t/(a/U_0)\geq100$,  and various order parameters are monitored to ensure steady states. To avoid metastability, we repeat the same set of simulations starting from dry MIPS configurations. 
We approximate the hard-sphere non-hydrodynamic force between particles with the WCA potential~\cite{weeks1971role}, which recovers the behavior of ABPs when not including HI.

\begin{figure}[thb]
\centering
\includegraphics[width=0.95\columnwidth]{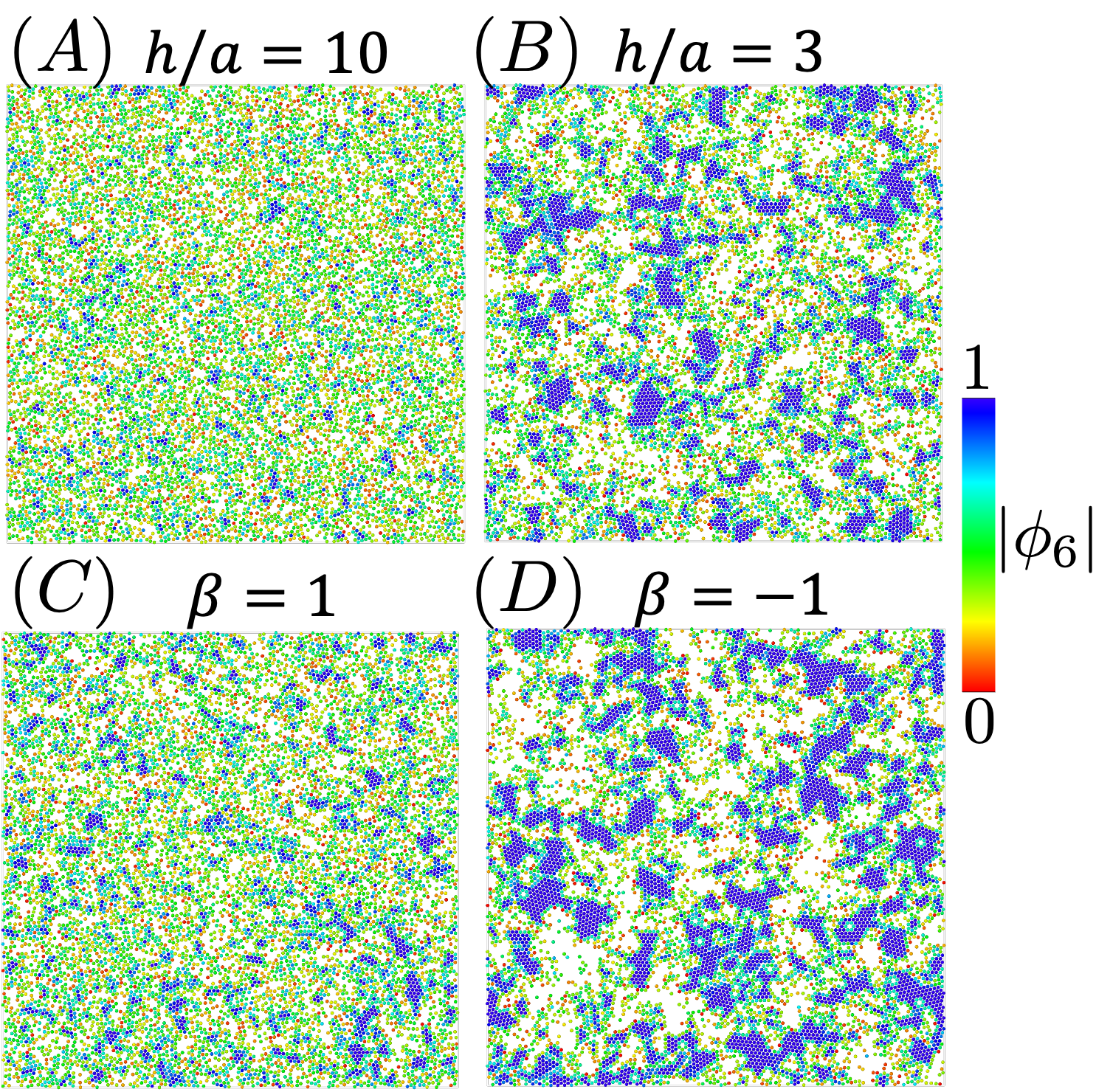}
\centering
\caption{Simulation snapshots at steady states for persistent swimmers ($Pe_R=0$) at $\phi=0.6$. Particles are colored by their local crystalline order parameter $|\phi_6|$.
(AB) Stealth swimmers $\beta=0$ with full HI at varying film thickness $h/a$.
(CD) ``pullers'' ($\beta>0$) and ``pushers'' ($\beta<0$) at $h/a=10.0$.
\label{fig:snapshots}}
\end{figure}

{\it Results---}
We vary the dimensionless reorientational Pecl\'et number, $Pe_R=a D_R/U_0 = a/\ell_0$, where $\ell_0$ is the free space run or persistence length, the packing density $\phi = n \pi a^2$, where $n$ is the areal number density, and the dimensionless film thickness $h/a$ systematically in the simulations. 
We observe that MIPS always disappears for stealth swimmers when full HI is included, as shown for the persistent swimmers ($Pe_R=0$) in Fig.~\ref{fig:snapshots}(AB).
We observe no MIPS for any combination of parameters ($Pe_R$, $h/a$, $\phi$).
Including the $B_2$ swimming mode -- pushers or pullers -- does not affect the results qualitatively, as shown in Fig.~\ref{fig:snapshots}(CD). The fragmented small clusters are only density fluctuations and they never coarsen, confirmed by simulations using a phase-separated configuration as the initial condition. The global cluster in dry MIPS dissolves into small fragments of similar sizes as seen here (see SI movie M1).
We also conducted fully 3D simulations (see SI movie M2) with full HI and they show no MIPS.

To understand how HI destroy MIPS,  we start by analyzing the simple problem of two stealth swimmers. A stealth swimmer, by definition, does not generate a flow when it is moving alone -- it behaves just like a dry ABP. Where then do HI come from for stealth swimmers?  Consider a head-on collision of two swimmers, as sketched in Fig.~\ref{fig:fig1}(C). Now both swimmers experience the excluded volume repulsive force from each other, and these inter-particle forces are conservative and obey Newton's third law ($\bF_1^p=-\bF_2^p$).  The force pair forms a pusher-like dipole of strength $2aF$ and generates a far-field fluid velocity that decays as $u\sim 1/r^2$ in 3D, or as $1/r$  when $r > h$ as the flow becomes two dimensional in the monolayer film. From Eq.~(\ref{eqn:EOM}), a distant thrid particle experiences a velocity $\bU_3 \sim (\boldsymbol{M}_{UF}^{32} - \boldsymbol{M}_{UF}^{31})\cdot \bF_2^p$, where $\boldsymbol{M}_{UF}$ is the mobility matrix coupling velocity to force ($\boldsymbol{\mathcal{M}}_{\cal FU} = \boldsymbol{\mathcal{R}}_{\cal FU}^{-1}$). Of course, colliding swimmers do not have to align perfectly, and these collision-induced dipoles are always present for any type of microswimmer, regardless of the higher-order details of the surface slip velocity and/or near-field interactions.

We analyze the effects of HI in more detail below, by separating the effects of HI-induced translation from those of rotation. Although it may not be easy to realize this in an experiment, it is possible to do so theoretically for spherical squirmers by assuming a suitable time-varying surface slip velocity for each squirmer.

\begin{figure}[hbt]
\centering
\includegraphics[width=0.95\columnwidth]{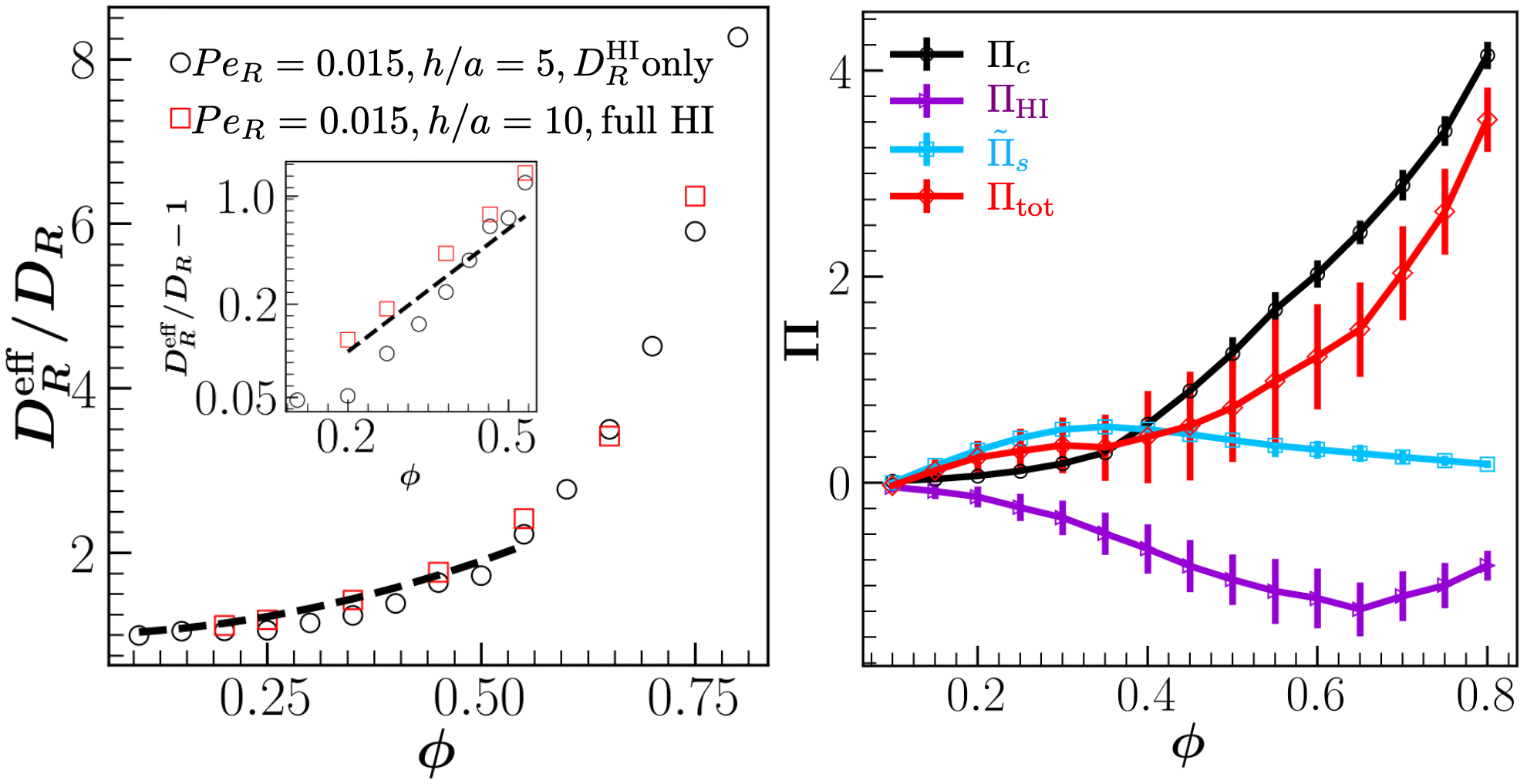}
\centering
\caption{
The effect of pure rotational HI. 
(A) $D_R^{\mathrm{eff}}$ scaling with packing density $\phi$ at $Pe_R=0.015$. Circles show simulation data for $h/a=5$ with only rotational HI, and squares show simulation data for $h/a=10$ with full HI. 
Black dash curve fits the scaling as in Eq.~(\ref{eq:DR-scaling}). The inset shows the scaling fit in a log-log scale.
(B) Pressure curves with vertical errorbars for $Pe_R=0.015$, $h/a=100$. Blue curve is the modified swim pressure constructed according to Eq.~(\ref{eq:modified-swim-pressure}). Black curve is the measured collisional pressure. Purple curve is the hydrodynamic pressure. Red curve is the total mechanical pressure. 
}
\label{fig:HIrotation}
\end{figure}

{\it Effects of pure rotational HI---}
First, we include only rotational HI, where the particle rotational velocities are updated from Eq.~(\ref{eqn:EOM}), but the translational velocities are not; in translation, the swimmer is a dry ABP with 
hard-sphere collisions without intervening HI. 
A collision-induced pusher dipole rotates a third particle via the vorticity of its flow field for the duration of this pair collision $\tau_C$. 
At the mean-field level, the randomly oriented collisional dipoles generate a fluctuating vorticity field that produces a hydrodynamic rotational diffusivity $D_R^{\mathrm{HI}}$ (see SI movie M3). This hydrodynamic diffusivity is in addition to the intrinsic rotational diffusivity $D_R$ of a swimmer.  
A simple calculation (See SI) shows that this HI contribution scales as $D_R^{\mathrm{HI}}\sim C_R(\phi)\phi^2(U_0/a)^2\tau_C$, where $C_R(\phi)>0$ approaches a constant as $\phi\rightarrow0$. The total, or effective, rotational diffusivity is $D_R^{\mathrm{eff}}=D_R+D_R^{\mathrm{HI}}$.  The collisional correlation time is estimated as $\tau_C\approx 1/D_R^{\mathrm{eff}}$, and thus
\begin{equation}
\frac{D_R^{\mathrm{eff}}}{D_R} = \frac{1 + \sqrt{1+4C_R \phi^2 \left(\frac{\ell_0}{a}\right)^2}}{2} \, .
\label{eq:DR-scaling}
\end{equation}
The formula above agrees reasonably well with our numerical data for $\phi\leq0.5$, as shown in Fig.~\ref{fig:HIrotation}(A). 

An enhanced $D_R^{\mathrm{eff}}$ means that particles spin so quickly that their effective run length $\ell^\mathrm{eff}  \sim U_0/D_R^{\mathrm{eff}} \ll \ell_0$ is reduced, and thus MIPS, which only occurs for very persistent swimmers ($\ell_0/a \gg 1$ or $Pe_R \ll 1$), cannot occur.  This can be most readily seen from a stability analysis of density fluctuations.  The particle number density $n$ satisfies 
\begin{equation}
\partial_t n = -\nabla\cdot (U_0\bmm - M\nabla\Pi_b),
\label{eq:mips-n}
\end{equation}
where $\bmm = \int P(\bx, \bq) d\bq$ is the polar order, $M$ is the average collective mobility in the suspension, and $\Pi_b=\Pi_c+\Pi_{\mathrm{HI}}$ is the bulk suspension  pressure. 
As the interaction between a pair of particles decomposes into the hard-sphere repulsion $\bF^P_{ij}$ and the HI force $\bF^{\mathrm{HI}}_{ij}$, $\Pi_b$ consists of the collisional,  $\Pi_c=\frac{1}{2V}\sum_{<i,j>}\br_{ij}\!\cdot\!\bF^p_{ij}$, and hydrodynamic,  $\Pi_{\mathrm{HI}}=\frac{1}{2V}\sum_{<i,j>}\br_{ij}\!\cdot\!\bF_{ij}^{\mathrm{HI}}$, pressures, respectively~\cite{nott1994pressure,wang2025reviving}. 

The polar order is governed by
\begin{equation}
    \partial_t \bmm  = -D_R^{\mathrm{eff}} \bmm - \nabla\cdot \left(\frac{U_0 \bar{U}n}{2}\bI + U_0 \bQ\right),
\end{equation}
where $\bQ$ is the nematic order tensor.  We keep only the first two moments, $n$ and $\bmm$, setting $\bQ = 0$, as they give the leading order behavior in the stability analysis~\cite{omar2023mechanical}.  Here, $\bar{U}\sim (1-n/n_{\mathrm{max}})$ is the reduced particle  velocity as the density increases~\cite{takatori2014swim}.

At steady state, we have $\bmm\approx-(1/D_R^{\mathrm{eff}})\nabla(U_0 \bar{U}n/2)$, from which we can define a modified `swim' pressure~\cite{takatori2014swim}
\begin{equation}
{\tilde{\Pi}_s}(n)=\frac{1}{M_0}\int_0^n dn \frac{1}{D_R^{\mathrm{eff}}(n)}\frac{d}{dn}\left(\frac{nU_0^2\bar{U}}{2}\right), 
\label{eq:modified-swim-pressure}
\end{equation}
which balances the suspension pressure  in Eq.~(\ref{eq:mips-n}). The isolated particle mobility $M_0$ is needed in order to have units of pressure. 
The collective mobility $M$ in Eq.~(\ref{eq:mips-n}) is a weakly decreasing function of denisty in the monolayer film, and thus we approximate it with $M_0$; the mobility then drops out of the analysis.

It is now straightforward to perform a linear stability analysis, and instability or phase separation happens when $\partial_n \Pi_{\mathrm{tot}}<0$, where $\Pi_{\mathrm{tot}}=\Pi_b+{\tilde{\Pi}}_s$. We construct ${\tilde{\Pi}}_s$ from the measured $D_R^{\mathrm{eff}}$. 
The total pressure $\Pi_{\mathrm{tot}}$ is always increasing monotonically, as shown in Fig.~\ref{fig:HIrotation}(B), in contrast to dry ABPs with the same parameters showing a clear plateau for phase separation. Hence, this simple analysis predicts no phase separation, confirmed in our simulations (SI movie M4).

Previous simulations of the `active disk' model~\cite{matas2014hydrodynamic}  found that the near-field lubrication interactions enhance the rotational diffusion so much that the effective persistence timescale is too short for MIPS to occur. Our results are consistent with this observation, although the enhancement of the rotational diffusion seen here comes primarily from the long-range collisional dipoles.

\begin{figure}[t]
\centering
\includegraphics[width=0.99\columnwidth]{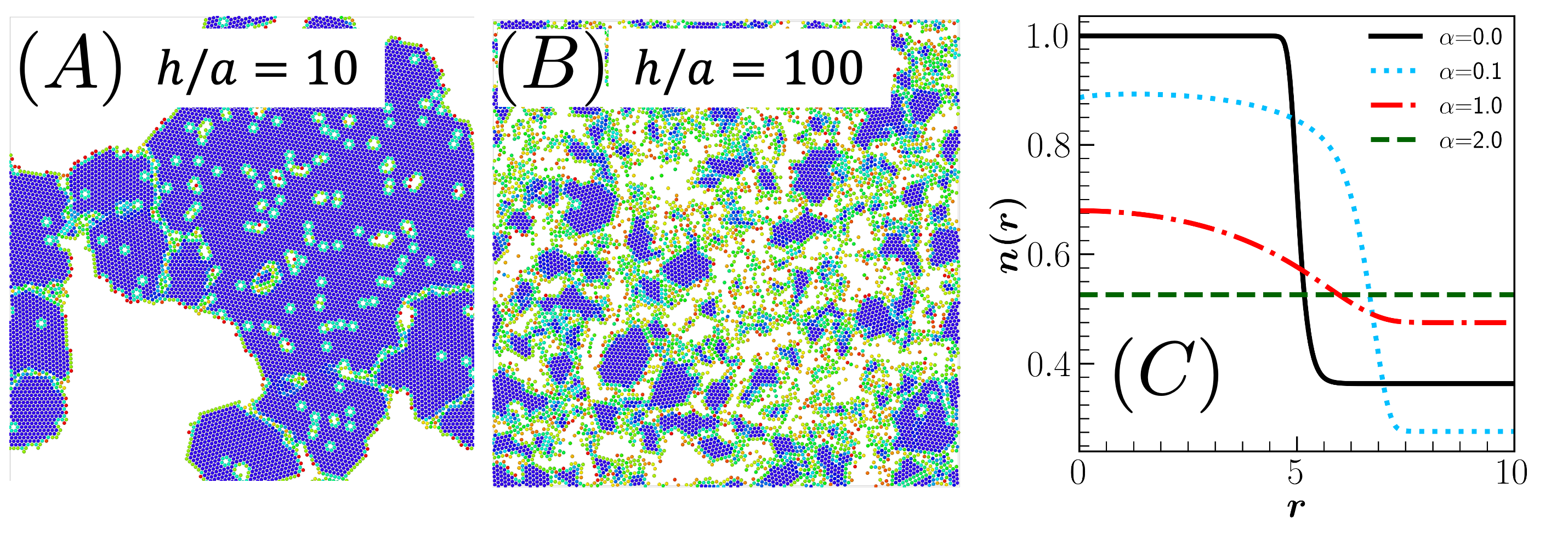}
\centering
\caption{
The effects of pure translational HI. 
(AB) Simulation snapshot for persistent swimmers at $\phi=0.6$, with (A) $h/a=10$ and (B) $h/a=100$.
(C) Steady state density profiles along the radial direction, solved from the local model Eq.~(\ref{eq:translationHI-nm}) assuming axisymmetry. The lines vary from black solid, blue dots, red dash-dots to green dashes as the flow strength parameter $\alpha$ increases. $\alpha=0$ (black solid) corresponds to dry MIPS. The y-axis is normalized by $n$ at maximum packing.
\label{fig:translationalHI}}
\end{figure}

{\it Effects of pure translational HI---}
While rotational HI always acts to prevent MIPS, the situation with only translational HI is more subtle. 
For translational HI only, the reorientation of swimmers is assumed to be entirely determined by their intrinsic noise and is not coupled to the flows.  We find that for thin films when $h \sim O(2a)$ and the systems is essentially 2D, MIPS occurs (Fig.~\ref{fig:translationalHI}A and SI movie M5), whereas for thick films $h \gg a$ MIPS disappears (Fig.~\ref{fig:translationalHI}B and SI movie M6).  Why?

Two different types of HI effects need to be considered. First, the flows generated by randomly oriented collisional dipoles contribute to an effective translational diffusivity $D_T^{\mathrm{HI}}\sim \bU_{\mathrm{HI}}^2\tau_C\sim C_T(\phi) \phi^2 U_0^2\tau_C$, similar to the argument for rotational diffusivity enhanced by HI.  However, $D_T^{\mathrm{HI}}$  enters the transport equations as a diffusive flux, $ - D_T^{\mathrm{HI}}\nabla n$, which is {\em added}  to other fluxes in Eq.~(\ref{eq:mips-n}). Consequently, the presence of $D_T^{\mathrm{HI}}$ only shifts the phase boundary of MIPS but will not forbid MIPS for persistent swimmers. 
Numerically, $D_T^{\mathrm{HI}}$ measured from simulations is always negligible compared to other fluxes and we will set $D_T^{\mathrm{HI}}=0$ going forward.

The second, more subtle, effect comes from the net advective flow of fluid (or more properly of suspension) produced by a distribution of collisional dipoles. From a continuum perspective, such flows at scales larger than particle size are caused by the gradient of the particle phase stress, $\nabla\Pi_c$, as we now show.  First, the transport equations for $n$ and $\bmm$ have an additional advective flux:
\begin{equation}
\begin{split}
\partial_t n & = -\nabla \cdot (U_0 \bmm -M \nabla\Pi_c + \boldsymbol{w} n),\\
\partial_t \bmm & = -D_R\bmm - \nabla\cdot \left(\frac{U_0 \bar{U}n}{2}\bI + \boldsymbol{w}\bmm \right),\\
\end{split}
\label{eq:translationHI-nm}
\end{equation}
where $\nabla=(\partial_x,\partial_y)$, and $\boldsymbol{w} =w_x\be_x+w_y\be_y$ is the 2D velocity field in the $xy$-plane.  The 3D suspension flow $\bu=(\boldsymbol{w},u_z)$ is found from a momentum balance on the suspension, where the stress has the usual viscous stress plus the collisional stress from the interparticle forces: $\boldsymbol{\sigma} = - (p + \Pi_c)\boldsymbol{I} + 2\mu \boldsymbol{e}$;
\begin{equation}
\begin{split}
\mu (\nabla^2+\partial_z^2)\bu & = (\nabla+\partial_z) (p + \Pi_c)\, ,\\
\nabla\cdot \bu & = 0.
\end{split}
\label{eq:flow-translationalHI}
\end{equation}
While the 3D flow $\bu$ is incompressible, its 2D projection $\boldsymbol{w} = \bu(x,y, z=0)$  need not be incompressible in the 2D plane, $\nabla\cdot\boldsymbol{w}\neq0$.

The flow is driven by $\nabla \Pi_c$, the latter of which is confined to the $xy$-plane.  If we have a dense MIPS-like cluster, there will be a rapid variation in number density as one crosses from the dilute region outside the cluster to its interior, and a corresponding gradient in collisional pressure that will tend to drive fluid motion away from the cluster.   If the film is thin, $h \sim O(2a)$, variations in $z$ are small and the fluid pressure field can balance the collisional pressure, $p \approx - \Pi_c$, such that  no flow occurs, $\bu \approx 0$.  
And since $D_T^{\mathrm{HI}}$ is too weak to stabilize the system, MIPS occurs as seen in Fig.~\ref{fig:translationalHI}(A).  The same is also true for fully 3D suspensions, where the fluid pressure balances the collisional pressure, there is no suspension flow, and MIPS occurs for persistent swimmers (SI movie M7).

For thicker films ($h \gg a$), however, the outward fluid flow produced at the edge of a cluster by $\nabla \Pi_c$ draws fluid in from above and below ($\pm z$) and this flow entrains swimmers and sweeps them away from the cluster and destroys MIPS (See SI Fig.1 and SI movie M8).
To get an estimate of this flow and to show how it impacts  MIPS, we consider the following model.

We replace $\nabla\Pi_c$ on the RHS of Eqn.~\ref{eq:flow-translationalHI} with point forces periodic in z-direction $-\bff\sum_n\delta(z-nh)$. Solving by Fourier transform gives $u_{xy,n}\sim K_0(\frac{2\pi}{n h}r)$ for modes $n\neq 0$, where $K_0(x)\sim\sqrt{x}e^{-x}$ is the modified Bessel function. 
($u_{xy,0}$ decays algebraically, but it does not contribute when the force density distribution is axisymmetric.) Hence, the leading order flow ($n=1$) decays on the scale of $h$. This motivates us to approximate it locally as $\bu(x,y, z=0) = \boldsymbol{w} \simeq -\alpha M_0\nabla\Pi_c$, where $\alpha\sim O(h/a)$ represents the strength of the flow and $M_0$ is the mobility.
Substituting this local approximation of $\boldsymbol{w}$ into Eq.~(\ref{eq:translationHI-nm}), we solve for the radial density profile at steady state, as shown in Fig.~\ref{fig:translationalHI}(C). 
This simple model recovers the dry MIPS when $\alpha=0$, and suggests MIPS to disappear when $\alpha\sim O(1)$. 
Hence, these model equations predict that phase separation is eventually suppressed when $h/a\sim\alpha\gtrsim O(1)$. 
To confirm this prediction, we conduct simulations of persistent swimmers where only the translational HI velocity is included, as shown in Fig.~\ref{fig:translationalHI}(AB). Indeed, MIPS is destroyed when $h\gg a$.

\begin{figure}[t]
\centering
\includegraphics[width=0.995\columnwidth]{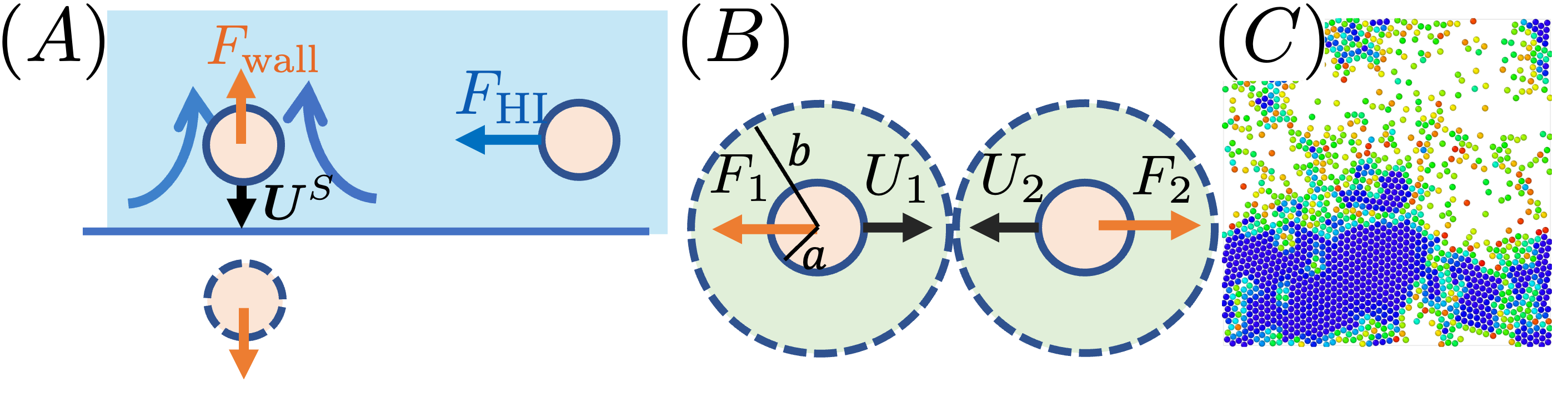}
\centering
\caption{
(A) Collision-induced dipole between a swimmer and wall generates an attractive flow.
(B) A larger radius $b$ for the excluded volume interaction than the radius $a$ for the particle-fluid surface, possibly by brush-like surface coating. 
(C) Simulation snapshot with full HI for $Pe_R=0$, $b/a=2$, $h/a=3$ at steady state. Particles are visualized with excluded volume radius $b$.
}
\label{fig:discussion}
\end{figure}

{\it Discussion---}
Several previous studies have observed particle aggregation and clustering in quasi-2D simulations~\cite{zottl2014hydrodynamics} or experiments~\cite{thutupalli2018flow}. However, as pointed out in \citep{thutupalli2018flow}, the attractive flow due to collisions of particles with the bottom wall is responsible for the aggregation, not active mobility itself. In fact, it was also noticed in \citep{zottl2014hydrodynamics} that the particles tend to orient towards one of the walls. Such attractive flow effect has appeared in other contexts~\cite{squires2000like,tan2022odd} as well, and can be modeled in the same framework as presented here of collisional dipoles, this time between particles and boundaries, as shown in Fig.~\ref{fig:discussion}(A). 

Finally, is it possible to mediate the strength of HI to recover MIPS in a suspension? One possibility is to introduce brush-like surface coatings on the particles, as shown in Fig.~\ref{fig:discussion}(B). The excluded volume interaction then acts with an effective particle radius $b$, larger than the radius $a$ that sets the scale for the HI~\cite{bergenholtz2002non}. 
The HI are now reduced by a factor of $(a/b)^2$, weakening the effect of rotational $D_R^{\mathrm{HI}}$. 
We confirm the reappearance of MIPS in these conditions with $b/a-1\gtrsim O(1)$ for full HI, as shown in Fig.~\ref{fig:discussion}(C) and SI movie 9.

{\it Conclusion---}
We provide theoretical and numerical results to show that MIPS is destroyed by the effects of full HI, even for the simplest squirmers, i.e., stealth swimmers. The fundamental mechanism is collision-induced dipoles among swimmers and between swimmers and boundaries, which are present for any type of microswimmer. We investigate separately the effects of rotational and translational HI, and show that both can independently destroy MIPS. Based on these results, we propose a method to control the magnitude of HI by surface coating (e.g. excluded volume interactions) to recover MIPS in a microfluidic experiment.

{\it Acknowledgments}---This work is financially supported by the Department of Energy (DOE) grant DE SCO22966.

\bibliography{literature}

\end{document}

%% file: pream.tex
\usepackage{graphicx}
\usepackage{cancel}
\usepackage{dcolumn}
\usepackage{bm}
\usepackage{siunitx}
\usepackage{xcolor}

\graphicspath{{figs/}}

\newcommand{\be}{\boldsymbol{e}}

\newcommand{\bff}{\boldsymbol{f}}
\newcommand{\bq}{\boldsymbol{q}}

\newcommand{\bmm}{\boldsymbol{m}}

\newcommand{\br}{\boldsymbol{r}}

\newcommand{\bx}{\boldsymbol{x}}

\newcommand{\bu}{\boldsymbol{u}}

\newcommand{\bU}{\boldsymbol{U}}

\newcommand{\bOmega}{\bm{\varOmega}}
\newcommand{\bF}{\boldsymbol{F}}
\newcommand{\bE}{\boldsymbol{E}}

\newcommand{\bI}{\boldsymbol{I}}

\newcommand{\bQ}{\boldsymbol{Q}}


%% file: main.bbl
\begin{thebibliography}{28}%
\makeatletter
\providecommand \@ifxundefined [1]{%
 \@ifx{#1\undefined}
}%
\providecommand \@ifnum [1]{%
 \ifnum #1\expandafter \@firstoftwo
 \else \expandafter \@secondoftwo
 \fi
}%
\providecommand \@ifx [1]{%
 \ifx #1\expandafter \@firstoftwo
 \else \expandafter \@secondoftwo
 \fi
}%
\providecommand \natexlab [1]{#1}%
\providecommand \enquote  [1]{``#1''}%
\providecommand \bibnamefont  [1]{#1}%
\providecommand \bibfnamefont [1]{#1}%
\providecommand \citenamefont [1]{#1}%
\providecommand \href@noop [0]{\@secondoftwo}%
\providecommand \href [0]{\begingroup \@sanitize@url \@href}%
\providecommand \@href[1]{\@@startlink{#1}\@@href}%
\providecommand \@@href[1]{\endgroup#1\@@endlink}%
\providecommand \@sanitize@url [0]{\catcode `\\12\catcode `\$12\catcode
  `\&12\catcode `\#12\catcode `\^12\catcode `\_12\catcode `\%12\relax}%
\providecommand \@@startlink[1]{}%
\providecommand \@@endlink[0]{}%
\providecommand \url  [0]{\begingroup\@sanitize@url \@url }%
\providecommand \@url [1]{\endgroup\@href {#1}{\urlprefix }}%
\providecommand \urlprefix  [0]{URL }%
\providecommand \Eprint [0]{\href }%
\providecommand \doibase [0]{http://dx.doi.org/}%
\providecommand \selectlanguage [0]{\@gobble}%
\providecommand \bibinfo  [0]{\@secondoftwo}%
\providecommand \bibfield  [0]{\@secondoftwo}%
\providecommand \translation [1]{[#1]}%
\providecommand \BibitemOpen [0]{}%
\providecommand \bibitemStop [0]{}%
\providecommand \bibitemNoStop [0]{.\EOS\space}%
\providecommand \EOS [0]{\spacefactor3000\relax}%
\providecommand \BibitemShut  [1]{\csname bibitem#1\endcsname}%
\let\auto@bib@innerbib\@empty
\bibitem [{\citenamefont {Dunkel}\ \emph {et~al.}(2013)\citenamefont {Dunkel},
  \citenamefont {Heidenreich}, \citenamefont {Drescher}, \citenamefont
  {Wensink}, \citenamefont {B{\"a}r},\ and\ \citenamefont
  {Goldstein}}]{dunkel2013fluid}%
  \BibitemOpen
  \bibfield  {author} {\bibinfo {author} {\bibfnamefont {J{\"o}rn}\
  \bibnamefont {Dunkel}}, \bibinfo {author} {\bibfnamefont {Sebastian}\
  \bibnamefont {Heidenreich}}, \bibinfo {author} {\bibfnamefont {Knut}\
  \bibnamefont {Drescher}}, \bibinfo {author} {\bibfnamefont {Henricus~H}\
  \bibnamefont {Wensink}}, \bibinfo {author} {\bibfnamefont {Markus}\
  \bibnamefont {B{\"a}r}}, \ and\ \bibinfo {author} {\bibfnamefont {Raymond~E}\
  \bibnamefont {Goldstein}},\ }\bibfield  {title} {\enquote {\bibinfo {title}
  {Fluid dynamics of bacterial turbulence},}\ }\href@noop {} {\bibfield
  {journal} {\bibinfo  {journal} {Physical Review Letters}\ }\textbf {\bibinfo
  {volume} {110}},\ \bibinfo {pages} {228102} (\bibinfo {year}
  {2013})}\BibitemShut {NoStop}%
\bibitem [{\citenamefont {Dombrowski}\ \emph {et~al.}(2004)\citenamefont
  {Dombrowski}, \citenamefont {Cisneros}, \citenamefont {Chatkaew},
  \citenamefont {Goldstein},\ and\ \citenamefont
  {Kessler}}]{dombrowski2004self}%
  \BibitemOpen
  \bibfield  {author} {\bibinfo {author} {\bibfnamefont {Christopher}\
  \bibnamefont {Dombrowski}}, \bibinfo {author} {\bibfnamefont {Luis}\
  \bibnamefont {Cisneros}}, \bibinfo {author} {\bibfnamefont {Sunita}\
  \bibnamefont {Chatkaew}}, \bibinfo {author} {\bibfnamefont {Raymond~E}\
  \bibnamefont {Goldstein}}, \ and\ \bibinfo {author} {\bibfnamefont {John~O}\
  \bibnamefont {Kessler}},\ }\bibfield  {title} {\enquote {\bibinfo {title}
  {Self-concentration and large-scale coherence in bacterial dynamics},}\
  }\href@noop {} {\bibfield  {journal} {\bibinfo  {journal} {Physical Review
  Letters}\ }\textbf {\bibinfo {volume} {93}},\ \bibinfo {pages} {098103}
  (\bibinfo {year} {2004})}\BibitemShut {NoStop}%
\bibitem [{\citenamefont {Cates}\ and\ \citenamefont
  {Tailleur}(2015)}]{cates2015motility}%
  \BibitemOpen
  \bibfield  {author} {\bibinfo {author} {\bibfnamefont {Michael~E}\
  \bibnamefont {Cates}}\ and\ \bibinfo {author} {\bibfnamefont {Julien}\
  \bibnamefont {Tailleur}},\ }\bibfield  {title} {\enquote {\bibinfo {title}
  {Motility-induced phase separation},}\ }\href@noop {} {\bibfield  {journal}
  {\bibinfo  {journal} {Annu. Rev. Condens. Matter Phys.}\ }\textbf {\bibinfo
  {volume} {6}},\ \bibinfo {pages} {219--244} (\bibinfo {year}
  {2015})}\BibitemShut {NoStop}%
\bibitem [{\citenamefont {Tailleur}\ and\ \citenamefont
  {Cates}(2008)}]{tailleur2008statistical}%
  \BibitemOpen
  \bibfield  {author} {\bibinfo {author} {\bibfnamefont {Julien}\ \bibnamefont
  {Tailleur}}\ and\ \bibinfo {author} {\bibfnamefont {Michael~E}\ \bibnamefont
  {Cates}},\ }\bibfield  {title} {\enquote {\bibinfo {title} {Statistical
  mechanics of interacting run-and-tumble bacteria},}\ }\href@noop {}
  {\bibfield  {journal} {\bibinfo  {journal} {Physical Review Letters}\
  }\textbf {\bibinfo {volume} {100}},\ \bibinfo {pages} {218103} (\bibinfo
  {year} {2008})}\BibitemShut {NoStop}%
\bibitem [{\citenamefont {Fily}\ and\ \citenamefont
  {Marchetti}(2012)}]{fily2012athermal}%
  \BibitemOpen
  \bibfield  {author} {\bibinfo {author} {\bibfnamefont {Yaouen}\ \bibnamefont
  {Fily}}\ and\ \bibinfo {author} {\bibfnamefont {M~Cristina}\ \bibnamefont
  {Marchetti}},\ }\bibfield  {title} {\enquote {\bibinfo {title} {Athermal
  phase separation of self-propelled particles with no alignment},}\
  }\href@noop {} {\bibfield  {journal} {\bibinfo  {journal} {Physical Review
  Letters}\ }\textbf {\bibinfo {volume} {108}},\ \bibinfo {pages} {235702}
  (\bibinfo {year} {2012})}\BibitemShut {NoStop}%
\bibitem [{\citenamefont {Speck}\ \emph {et~al.}(2014)\citenamefont {Speck},
  \citenamefont {Bialk{\'e}}, \citenamefont {Menzel},\ and\ \citenamefont
  {L{\"o}wen}}]{speck2014effective}%
  \BibitemOpen
  \bibfield  {author} {\bibinfo {author} {\bibfnamefont {Thomas}\ \bibnamefont
  {Speck}}, \bibinfo {author} {\bibfnamefont {Julian}\ \bibnamefont
  {Bialk{\'e}}}, \bibinfo {author} {\bibfnamefont {Andreas~M}\ \bibnamefont
  {Menzel}}, \ and\ \bibinfo {author} {\bibfnamefont {Hartmut}\ \bibnamefont
  {L{\"o}wen}},\ }\bibfield  {title} {\enquote {\bibinfo {title} {Effective
  cahn-hilliard equation for the phase separation of active brownian
  particles},}\ }\href@noop {} {\bibfield  {journal} {\bibinfo  {journal}
  {Physical Review Letters}\ }\textbf {\bibinfo {volume} {112}},\ \bibinfo
  {pages} {218304} (\bibinfo {year} {2014})}\BibitemShut {NoStop}%
\bibitem [{\citenamefont {Takatori}\ \emph {et~al.}(2014)\citenamefont
  {Takatori}, \citenamefont {Yan},\ and\ \citenamefont
  {Brady}}]{takatori2014swim}%
  \BibitemOpen
  \bibfield  {author} {\bibinfo {author} {\bibfnamefont {Sho~C}\ \bibnamefont
  {Takatori}}, \bibinfo {author} {\bibfnamefont {Wen}\ \bibnamefont {Yan}}, \
  and\ \bibinfo {author} {\bibfnamefont {John~F}\ \bibnamefont {Brady}},\
  }\bibfield  {title} {\enquote {\bibinfo {title} {Swim pressure: stress
  generation in active matter},}\ }\href@noop {} {\bibfield  {journal}
  {\bibinfo  {journal} {Physical Review Letters}\ }\textbf {\bibinfo {volume}
  {113}},\ \bibinfo {pages} {028103} (\bibinfo {year} {2014})}\BibitemShut
  {NoStop}%
\bibitem [{\citenamefont {Omar}\ \emph {et~al.}(2023)\citenamefont {Omar},
  \citenamefont {Row}, \citenamefont {Mallory},\ and\ \citenamefont
  {Brady}}]{omar2023mechanical}%
  \BibitemOpen
  \bibfield  {author} {\bibinfo {author} {\bibfnamefont {Ahmad~K}\ \bibnamefont
  {Omar}}, \bibinfo {author} {\bibfnamefont {Hyeongjoo}\ \bibnamefont {Row}},
  \bibinfo {author} {\bibfnamefont {Stewart~A}\ \bibnamefont {Mallory}}, \ and\
  \bibinfo {author} {\bibfnamefont {John~F}\ \bibnamefont {Brady}},\ }\bibfield
   {title} {\enquote {\bibinfo {title} {Mechanical theory of nonequilibrium
  coexistence and motility-induced phase separation},}\ }\href@noop {}
  {\bibfield  {journal} {\bibinfo  {journal} {Proceedings of the National
  Academy of Sciences}\ }\textbf {\bibinfo {volume} {120}},\ \bibinfo {pages}
  {e2219900120} (\bibinfo {year} {2023})}\BibitemShut {NoStop}%
\bibitem [{\citenamefont {Lighthill}(1952)}]{lighthill1952squirming}%
  \BibitemOpen
  \bibfield  {author} {\bibinfo {author} {\bibfnamefont {Michael~James}\
  \bibnamefont {Lighthill}},\ }\bibfield  {title} {\enquote {\bibinfo {title}
  {On the squirming motion of nearly spherical deformable bodies through
  liquids at very small reynolds numbers},}\ }\href@noop {} {\bibfield
  {journal} {\bibinfo  {journal} {Communications on Pure and Applied
  Mathematics}\ }\textbf {\bibinfo {volume} {5}},\ \bibinfo {pages} {109--118}
  (\bibinfo {year} {1952})}\BibitemShut {NoStop}%
\bibitem [{\citenamefont {Blake}(1971)}]{blake1971spherical}%
  \BibitemOpen
  \bibfield  {author} {\bibinfo {author} {\bibfnamefont {John~R}\ \bibnamefont
  {Blake}},\ }\bibfield  {title} {\enquote {\bibinfo {title} {A spherical
  envelope approach to ciliary propulsion},}\ }\href@noop {} {\bibfield
  {journal} {\bibinfo  {journal} {Journal of Fluid Mechanics}\ }\textbf
  {\bibinfo {volume} {46}},\ \bibinfo {pages} {199--208} (\bibinfo {year}
  {1971})}\BibitemShut {NoStop}%
\bibitem [{\citenamefont {Saintillan}\ and\ \citenamefont
  {Shelley}(2008)}]{saintillan2008instabilities}%
  \BibitemOpen
  \bibfield  {author} {\bibinfo {author} {\bibfnamefont {David}\ \bibnamefont
  {Saintillan}}\ and\ \bibinfo {author} {\bibfnamefont {Michael~J}\
  \bibnamefont {Shelley}},\ }\bibfield  {title} {\enquote {\bibinfo {title}
  {Instabilities, pattern formation, and mixing in active suspensions},}\
  }\href@noop {} {\bibfield  {journal} {\bibinfo  {journal} {Physics of
  Fluids}\ }\textbf {\bibinfo {volume} {20}} (\bibinfo {year}
  {2008})}\BibitemShut {NoStop}%
\bibitem [{\citenamefont {Ishikawa}\ and\ \citenamefont
  {Pedley}(2008)}]{ishikawa2008coherent}%
  \BibitemOpen
  \bibfield  {author} {\bibinfo {author} {\bibfnamefont {Takuji}\ \bibnamefont
  {Ishikawa}}\ and\ \bibinfo {author} {\bibfnamefont {Timothy~J}\ \bibnamefont
  {Pedley}},\ }\bibfield  {title} {\enquote {\bibinfo {title} {Coherent
  structures in monolayers of swimming particles},}\ }\href@noop {} {\bibfield
  {journal} {\bibinfo  {journal} {Physical Review Letters}\ }\textbf {\bibinfo
  {volume} {100}},\ \bibinfo {pages} {088103} (\bibinfo {year}
  {2008})}\BibitemShut {NoStop}%
\bibitem [{\citenamefont {Ishikawa}\ \emph {et~al.}(2008)\citenamefont
  {Ishikawa}, \citenamefont {Locsei},\ and\ \citenamefont
  {Pedley}}]{ishikawa2008development}%
  \BibitemOpen
  \bibfield  {author} {\bibinfo {author} {\bibfnamefont {Takuji}\ \bibnamefont
  {Ishikawa}}, \bibinfo {author} {\bibfnamefont {JT}~\bibnamefont {Locsei}}, \
  and\ \bibinfo {author} {\bibfnamefont {TJ}~\bibnamefont {Pedley}},\
  }\bibfield  {title} {\enquote {\bibinfo {title} {Development of coherent
  structures in concentrated suspensions of swimming model micro-organisms},}\
  }\href@noop {} {\bibfield  {journal} {\bibinfo  {journal} {Journal of Fluid
  Mechanics}\ }\textbf {\bibinfo {volume} {615}},\ \bibinfo {pages} {401--431}
  (\bibinfo {year} {2008})}\BibitemShut {NoStop}%
\bibitem [{\citenamefont {Drescher}\ \emph {et~al.}(2009)\citenamefont
  {Drescher}, \citenamefont {Leptos}, \citenamefont {Tuval}, \citenamefont
  {Ishikawa}, \citenamefont {Pedley},\ and\ \citenamefont
  {Goldstein}}]{drescher2009dancing}%
  \BibitemOpen
  \bibfield  {author} {\bibinfo {author} {\bibfnamefont {Knut}\ \bibnamefont
  {Drescher}}, \bibinfo {author} {\bibfnamefont {Kyriacos~C}\ \bibnamefont
  {Leptos}}, \bibinfo {author} {\bibfnamefont {Idan}\ \bibnamefont {Tuval}},
  \bibinfo {author} {\bibfnamefont {Takuji}\ \bibnamefont {Ishikawa}}, \bibinfo
  {author} {\bibfnamefont {Timothy~J}\ \bibnamefont {Pedley}}, \ and\ \bibinfo
  {author} {\bibfnamefont {Raymond~E}\ \bibnamefont {Goldstein}},\ }\bibfield
  {title} {\enquote {\bibinfo {title} {Dancing volvox: hydrodynamic bound
  states of swimming algae},}\ }\href@noop {} {\bibfield  {journal} {\bibinfo
  {journal} {Physical Review Letters}\ }\textbf {\bibinfo {volume} {102}},\
  \bibinfo {pages} {168101} (\bibinfo {year} {2009})}\BibitemShut {NoStop}%
\bibitem [{\citenamefont {Ishikawa}\ \emph {et~al.}(2020)\citenamefont
  {Ishikawa}, \citenamefont {Pedley}, \citenamefont {Drescher},\ and\
  \citenamefont {Goldstein}}]{ishikawa2020stability}%
  \BibitemOpen
  \bibfield  {author} {\bibinfo {author} {\bibfnamefont {Takuji}\ \bibnamefont
  {Ishikawa}}, \bibinfo {author} {\bibfnamefont {TJ}~\bibnamefont {Pedley}},
  \bibinfo {author} {\bibfnamefont {Knut}\ \bibnamefont {Drescher}}, \ and\
  \bibinfo {author} {\bibfnamefont {Raymond~E}\ \bibnamefont {Goldstein}},\
  }\bibfield  {title} {\enquote {\bibinfo {title} {Stability of dancing
  volvox},}\ }\href@noop {} {\bibfield  {journal} {\bibinfo  {journal} {Journal
  of Fluid Mechanics}\ }\textbf {\bibinfo {volume} {903}},\ \bibinfo {pages}
  {A11} (\bibinfo {year} {2020})}\BibitemShut {NoStop}%
\bibitem [{\citenamefont {Matas-Navarro}\ \emph {et~al.}(2014)\citenamefont
  {Matas-Navarro}, \citenamefont {Golestanian}, \citenamefont {Liverpool},\
  and\ \citenamefont {Fielding}}]{matas2014hydrodynamic}%
  \BibitemOpen
  \bibfield  {author} {\bibinfo {author} {\bibfnamefont {Ricard}\ \bibnamefont
  {Matas-Navarro}}, \bibinfo {author} {\bibfnamefont {Ramin}\ \bibnamefont
  {Golestanian}}, \bibinfo {author} {\bibfnamefont {Tanniemola~B}\ \bibnamefont
  {Liverpool}}, \ and\ \bibinfo {author} {\bibfnamefont {Suzanne~M}\
  \bibnamefont {Fielding}},\ }\bibfield  {title} {\enquote {\bibinfo {title}
  {Hydrodynamic suppression of phase separation in active suspensions},}\
  }\href@noop {} {\bibfield  {journal} {\bibinfo  {journal} {Physical Review
  E}\ }\textbf {\bibinfo {volume} {90}},\ \bibinfo {pages} {032304} (\bibinfo
  {year} {2014})}\BibitemShut {NoStop}%
\bibitem [{\citenamefont {Z{\"o}ttl}\ and\ \citenamefont
  {Stark}(2014)}]{zottl2014hydrodynamics}%
  \BibitemOpen
  \bibfield  {author} {\bibinfo {author} {\bibfnamefont {Andreas}\ \bibnamefont
  {Z{\"o}ttl}}\ and\ \bibinfo {author} {\bibfnamefont {Holger}\ \bibnamefont
  {Stark}},\ }\bibfield  {title} {\enquote {\bibinfo {title} {Hydrodynamics
  determines collective motion and phase behavior of active colloids in
  quasi-two-dimensional confinement},}\ }\href@noop {} {\bibfield  {journal}
  {\bibinfo  {journal} {Physical Review Letters}\ }\textbf {\bibinfo {volume}
  {112}},\ \bibinfo {pages} {118101} (\bibinfo {year} {2014})}\BibitemShut
  {NoStop}%
\bibitem [{\citenamefont {Squires}\ and\ \citenamefont
  {Brenner}(2000)}]{squires2000like}%
  \BibitemOpen
  \bibfield  {author} {\bibinfo {author} {\bibfnamefont {Todd~M}\ \bibnamefont
  {Squires}}\ and\ \bibinfo {author} {\bibfnamefont {Michael~P}\ \bibnamefont
  {Brenner}},\ }\bibfield  {title} {\enquote {\bibinfo {title} {Like-charge
  attraction and hydrodynamic interaction},}\ }\href@noop {} {\bibfield
  {journal} {\bibinfo  {journal} {Physical Review Letters}\ }\textbf {\bibinfo
  {volume} {85}},\ \bibinfo {pages} {4976} (\bibinfo {year}
  {2000})}\BibitemShut {NoStop}%
\bibitem [{\citenamefont {Brady}\ and\ \citenamefont
  {Bossis}(1988)}]{brady1988stokesian}%
  \BibitemOpen
  \bibfield  {author} {\bibinfo {author} {\bibfnamefont {John~F}\ \bibnamefont
  {Brady}}\ and\ \bibinfo {author} {\bibfnamefont {Georges}\ \bibnamefont
  {Bossis}},\ }\bibfield  {title} {\enquote {\bibinfo {title} {Stokesian
  dynamics},}\ }\href@noop {} {\bibfield  {journal} {\bibinfo  {journal}
  {Annual Review of Fluid Mechanics}\ }\textbf {\bibinfo {volume} {20}},\
  \bibinfo {pages} {111--157} (\bibinfo {year} {1988})}\BibitemShut {NoStop}%
\bibitem [{\citenamefont {Fiore}\ and\ \citenamefont
  {Swan}(2019)}]{fiore2019fast}%
  \BibitemOpen
  \bibfield  {author} {\bibinfo {author} {\bibfnamefont {Andrew~M}\
  \bibnamefont {Fiore}}\ and\ \bibinfo {author} {\bibfnamefont {James~W}\
  \bibnamefont {Swan}},\ }\bibfield  {title} {\enquote {\bibinfo {title} {Fast
  stokesian dynamics},}\ }\href@noop {} {\bibfield  {journal} {\bibinfo
  {journal} {Journal of Fluid Mechanics}\ }\textbf {\bibinfo {volume} {878}},\
  \bibinfo {pages} {544--597} (\bibinfo {year} {2019})}\BibitemShut {NoStop}%
\bibitem [{\citenamefont {Elfring}\ and\ \citenamefont
  {Brady}(2022)}]{elfring2022active}%
  \BibitemOpen
  \bibfield  {author} {\bibinfo {author} {\bibfnamefont {Gwynn~J}\ \bibnamefont
  {Elfring}}\ and\ \bibinfo {author} {\bibfnamefont {John~F}\ \bibnamefont
  {Brady}},\ }\bibfield  {title} {\enquote {\bibinfo {title} {Active stokesian
  dynamics},}\ }\href@noop {} {\bibfield  {journal} {\bibinfo  {journal}
  {Journal of Fluid Mechanics}\ }\textbf {\bibinfo {volume} {952}},\ \bibinfo
  {pages} {A19} (\bibinfo {year} {2022})}\BibitemShut {NoStop}%
\bibitem [{\citenamefont {Lorentz}(1907)}]{lorentz1907allgemeiner}%
  \BibitemOpen
  \bibfield  {author} {\bibinfo {author} {\bibfnamefont {Hendrik~Antoon}\
  \bibnamefont {Lorentz}},\ }\bibfield  {title} {\enquote {\bibinfo {title}
  {Ein allgemeiner satz, die bewegung einer reibenden fl{\"u}ssigkeit
  betreffend, nebst einigen anwendungen desselben},}\ }\href@noop {} {\bibfield
   {journal} {\bibinfo  {journal} {Abh. Theor. Phys}\ }\textbf {\bibinfo
  {volume} {1}},\ \bibinfo {pages} {23} (\bibinfo {year} {1907})}\BibitemShut
  {NoStop}%
\bibitem [{\citenamefont {Weeks}\ \emph {et~al.}(1971)\citenamefont {Weeks},
  \citenamefont {Chandler},\ and\ \citenamefont {Andersen}}]{weeks1971role}%
  \BibitemOpen
  \bibfield  {author} {\bibinfo {author} {\bibfnamefont {John~D}\ \bibnamefont
  {Weeks}}, \bibinfo {author} {\bibfnamefont {David}\ \bibnamefont {Chandler}},
  \ and\ \bibinfo {author} {\bibfnamefont {Hans~C}\ \bibnamefont {Andersen}},\
  }\bibfield  {title} {\enquote {\bibinfo {title} {Role of repulsive forces in
  determining the equilibrium structure of simple liquids},}\ }\href@noop {}
  {\bibfield  {journal} {\bibinfo  {journal} {The Journal of Chemical Physics}\
  }\textbf {\bibinfo {volume} {54}},\ \bibinfo {pages} {5237--5247} (\bibinfo
  {year} {1971})}\BibitemShut {NoStop}%
\bibitem [{\citenamefont {Nott}\ and\ \citenamefont
  {Brady}(1994)}]{nott1994pressure}%
  \BibitemOpen
  \bibfield  {author} {\bibinfo {author} {\bibfnamefont {Prabhu~R}\
  \bibnamefont {Nott}}\ and\ \bibinfo {author} {\bibfnamefont {John~F}\
  \bibnamefont {Brady}},\ }\bibfield  {title} {\enquote {\bibinfo {title}
  {Pressure-driven flow of suspensions: simulation and theory},}\ }\href@noop
  {} {\bibfield  {journal} {\bibinfo  {journal} {Journal of Fluid Mechanics}\
  }\textbf {\bibinfo {volume} {275}},\ \bibinfo {pages} {157--199} (\bibinfo
  {year} {1994})}\BibitemShut {NoStop}%
\bibitem [{\citenamefont {Wang}\ \emph {et~al.}()\citenamefont {Wang},
  \citenamefont {Zhou},\ and\ \citenamefont {Brady}}]{wang2025reviving}%
  \BibitemOpen
  \bibfield  {author} {\bibinfo {author} {\bibfnamefont {Mu}~\bibnamefont
  {Wang}}, \bibinfo {author} {\bibfnamefont {Tingtao}\ \bibnamefont {Zhou}}, \
  and\ \bibinfo {author} {\bibfnamefont {John~F.}\ \bibnamefont {Brady}},\
  }\bibfield  {title} {\enquote {\bibinfo {title} {Reviving the suspension
  balance model},}\ }\href@noop {} {\bibinfo  {journal} {submitted to Physical
  Review Fluids.}\ }\BibitemShut {NoStop}%
\bibitem [{\citenamefont {Thutupalli}\ \emph {et~al.}(2018)\citenamefont
  {Thutupalli}, \citenamefont {Geyer}, \citenamefont {Singh}, \citenamefont
  {Adhikari},\ and\ \citenamefont {Stone}}]{thutupalli2018flow}%
  \BibitemOpen
\bibfield  {journal} {  }\bibfield  {author} {\bibinfo {author} {\bibfnamefont
  {Shashi}\ \bibnamefont {Thutupalli}}, \bibinfo {author} {\bibfnamefont
  {Delphine}\ \bibnamefont {Geyer}}, \bibinfo {author} {\bibfnamefont {Rajesh}\
  \bibnamefont {Singh}}, \bibinfo {author} {\bibfnamefont {Ronojoy}\
  \bibnamefont {Adhikari}}, \ and\ \bibinfo {author} {\bibfnamefont {Howard~A}\
  \bibnamefont {Stone}},\ }\bibfield  {title} {\enquote {\bibinfo {title}
  {Flow-induced phase separation of active particles is controlled by boundary
  conditions},}\ }\href@noop {} {\bibfield  {journal} {\bibinfo  {journal}
  {Proceedings of the National Academy of Sciences}\ }\textbf {\bibinfo
  {volume} {115}},\ \bibinfo {pages} {5403--5408} (\bibinfo {year}
  {2018})}\BibitemShut {NoStop}%
\bibitem [{\citenamefont {Tan}\ \emph {et~al.}(2022)\citenamefont {Tan},
  \citenamefont {Mietke}, \citenamefont {Li}, \citenamefont {Chen},
  \citenamefont {Higinbotham}, \citenamefont {Foster}, \citenamefont {Gokhale},
  \citenamefont {Dunkel},\ and\ \citenamefont {Fakhri}}]{tan2022odd}%
  \BibitemOpen
  \bibfield  {author} {\bibinfo {author} {\bibfnamefont {Tzer~Han}\
  \bibnamefont {Tan}}, \bibinfo {author} {\bibfnamefont {Alexander}\
  \bibnamefont {Mietke}}, \bibinfo {author} {\bibfnamefont {Junang}\
  \bibnamefont {Li}}, \bibinfo {author} {\bibfnamefont {Yuchao}\ \bibnamefont
  {Chen}}, \bibinfo {author} {\bibfnamefont {Hugh}\ \bibnamefont
  {Higinbotham}}, \bibinfo {author} {\bibfnamefont {Peter~J}\ \bibnamefont
  {Foster}}, \bibinfo {author} {\bibfnamefont {Shreyas}\ \bibnamefont
  {Gokhale}}, \bibinfo {author} {\bibfnamefont {J{\"o}rn}\ \bibnamefont
  {Dunkel}}, \ and\ \bibinfo {author} {\bibfnamefont {Nikta}\ \bibnamefont
  {Fakhri}},\ }\bibfield  {title} {\enquote {\bibinfo {title} {Odd dynamics of
  living chiral crystals},}\ }\href@noop {} {\bibfield  {journal} {\bibinfo
  {journal} {Nature}\ }\textbf {\bibinfo {volume} {607}},\ \bibinfo {pages}
  {287--293} (\bibinfo {year} {2022})}\BibitemShut {NoStop}%
\bibitem [{\citenamefont {Bergenholtz}\ \emph {et~al.}(2002)\citenamefont
  {Bergenholtz}, \citenamefont {Brady},\ and\ \citenamefont
  {Vicic}}]{bergenholtz2002non}%
  \BibitemOpen
  \bibfield  {author} {\bibinfo {author} {\bibfnamefont {Johan}\ \bibnamefont
  {Bergenholtz}}, \bibinfo {author} {\bibfnamefont {John~F.}\ \bibnamefont
  {Brady}}, \ and\ \bibinfo {author} {\bibfnamefont {Michael}\ \bibnamefont
  {Vicic}},\ }\bibfield  {title} {\enquote {\bibinfo {title} {The non-newtonian
  rheology of dilute colloidal suspensions},}\ }\href@noop {} {\bibfield
  {journal} {\bibinfo  {journal} {Journal of Fluid Mechanics}\ }\textbf
  {\bibinfo {volume} {456}},\ \bibinfo {pages} {239--275} (\bibinfo {year}
  {2002})}\BibitemShut {NoStop}%
\end{thebibliography}%
